\documentclass[%twocolumn,%showpacs,preprintnumbers,
amsmath,%amssymb,aps,
prd,nofootinbib,floatfix,11pt,%preprint,
]{revtex4}

\usepackage{hyperref}
\usepackage{placeins}
\usepackage{afterpage}

\usepackage{graphicx}% Include figure filesf
\usepackage{setspace}
\usepackage{bm}% bold math
\usepackage[dvips]{color}
\definecolor{Red}{cmyk}{0,1,1,0}
\definecolor{BrickRed}{cmyk}{0,0.89,0.94,0.28}
\definecolor{Blue}{cmyk}{1,1,0,0}
\definecolor{Green}{cmyk}{1,0,1,0}

\newcommand\beq{\begin{eqnarray}}
\newcommand\eeq{\end{eqnarray}}

\def\lsim{\mathrel{\rlap{\lower4pt\hbox{$\sim$}}
    \raise1pt\hbox{$<$}}}                % less than or approx. symbol
\def\gsim{\mathrel{\rlap{\lower4pt\hbox{$\sim$}}
    \raise1pt\hbox{$>$}}}            

\allowdisplaybreaks
\newcommand\lnbar{\overline{\ln}}
\newcommand\MSbar{$\overline{\rm{MS}}$ }

\begin{document}

\renewcommand{\theequation}{\arabic{section}.\arabic{equation}}
\renewcommand{\thefigure}{\arabic{section}.\arabic{figure}}
\renewcommand{\thetable}{\arabic{section}.\arabic{table}}

\title{\Large \baselineskip=20pt 
Three-loop corrections to the Fermi decay constant\\ in the $\overline{\rm{MS}}$ scheme}
\author{Stephen P.~Martin}
\affiliation{\mbox{\it Department of Physics, Northern Illinois University, DeKalb IL 60115}}

\begin{abstract}\normalsize \baselineskip=16pt
I present the leading 3-loop contributions to the Fermi decay constant in the Standard Model, using the tadpole-free pure $\overline{\rm{MS}}$ scheme. The calculation is exact in the limit in which the QCD coupling, the top-quark Yukawa coupling, and the square root of the Higgs self-coupling are all treated as large compared to the electroweak gauge couplings. The effect of the 3-loop contribution is to decrease the estimate of the Higgs VEV, for fixed on-shell inputs, by about 3.5 MeV. The renormalization scale dependence of the computed Fermi constant is greatly reduced compared to the previously known complete 2-loop order result, and as a fraction is now less than $\pm 2 \times 10^{-6}$ for renormalization scales between 90 and 250 GeV. This corresponds to a theoretical uncertainty in the VEV that is well under 1 MeV, and much smaller than the current parametric uncertainties coming mostly from the top-quark mass. I also comment on the impact on the precise prediction of the $W$-boson mass.
\end{abstract}

\maketitle

%\tableofcontents

\baselineskip=15.7pt

%\newpage

%%%%%%%%%%%%%%%%%%%%%%%%%%%%%%%%%%%%%%%%%%%%%%%%%%%%%%%%%%%%%%%
\section{Introduction\label{sec:intro}}
\setcounter{equation}{0}
\setcounter{figure}{0}
\setcounter{table}{0} 
\setcounter{footnote}{1}

In the Standard Model of particle physics, the calculation for the inverse lifetime of the muon can be factorized as
\beq
\frac{1}{\tau_\mu} &=&
\frac{G_F^2 M_{\mu}^5}{192 \pi^3}\hspace{0.8pt} F_{\rm kin}(M_e^2/M_\mu^2) 
\left (1 + \delta_{\rm QED} \right ) 
\left (1 + \frac{3}{5} \frac{M_\mu^2}{M_W^2} \right ),
\label{eq:GFfromtaumu}
\eeq
where $F_{\rm kin}(x) = 1 - 8 x + 8 x^3 - x^4 - 12 x^2 \ln(x)$ is the 3-body decay phase-space kinematics factor, and $\delta_{\rm QED}$ consists of the QED (with associated hadronic polarization effects) radiative corrections  in the low-energy effective theory, which have been obtained at 1-, 2-, and 3-loop orders, respectively, in 
\cite{Kinoshita:1958ru,Nir:1989rm},
\cite{vanRitbergen:1999fi,Steinhauser:1999bx,Pak:2008qt}, and
\cite{Ferroglia:1999tg,Fael:2020tow,Czakon:2021ybq}. The present experimental value \cite{MuLan:2010shf,MuLan:2012sih,ParticleDataGroup:2024cfk} of the 
Fermi constant defined in this way is\footnote{Recent versions of the Particle Data Group's {\em Review of particle physics} 
\cite{ParticleDataGroup:2024cfk} choose to absorb the last factor in eq.~(\ref{eq:GFfromtaumu}) into the definition of $G_F$, despite the fact that it is specific to muon decay. The $G_F$ as defined here by eq.~(\ref{eq:GFfromtaumu}) and quoted in eq.~(\ref{eq:GFexp}) is therefore smaller than that in ref.~\cite{ParticleDataGroup:2024cfk} by a fractional amount $(3/10) M_\mu^2/M_W^2 = 5.2 \times 10^{-7}$. This difference happens to be close to the present experimental uncertainty in $G_F$.}
\beq
G_F &=& 1.1663782(6) \times 10^{-5} \>{\rm GeV}^{-2} .
\label{eq:GFexp}
\eeq

The Fermi constant can be related to other parameters in the Standard Model in several different ways. A tremendous amount of effort has gone into evaluating the radiative corrections to $G_F$, motivated in part by the fact that this allows a prediction of the $W$-boson mass in terms of other quantities. Historically, the most popular way of organizing these results is in terms
of the on-shell scheme expression
\beq
G_F &=& \frac{\pi\alpha}{\sqrt{2} M_W^2 (1 - M_W^2/M_Z^2)} (1 + \Delta r)
\label{eq:defineDeltar}
\eeq
where $\alpha = 1/137.035999178\ldots$ is the fine-structure constant for low-energy Thomson scattering, and $M_W$ and $M_Z$ are the on-shell masses of the weak vector bosons. This defines a calculable quantity $\Delta r$. Because the low-energy QED corrections parameterized by $\delta_{\rm QED}$ have been factored out, $\Delta r$ can be obtained in terms of Feynman diagrams involving only high-energy effects.
It has been calculated completely in the Standard Model at 1-loop \cite{Sirlin:1980nh,Marciano:1980pb,Sirlin:1981yz}
and 2-loop \cite{vanderBij:1986hy,Djouadi:1987gn,Djouadi:1987di,Kniehl:1989yc,Halzen:1990je,Barbieri:1992nz,Barbieri:1992dq,Fleischer:1993ub,Djouadi:1993ss,Degrassi:1996mg,Degrassi:1996ps,Malde:1999bd,Freitas:2000gg,Freitas:2002ja,Awramik:2002wn,Onishchenko:2002ve,Awramik:2002vu,Awramik:2003ee}
orders. In addition, there are partial 3-loop contributions to the $\rho$ parameter, of orders $g_3^4 y_t^2$ \cite{Avdeev:1994db,Chetyrkin:1995ix,Chetyrkin:1995js},
and $g_3^2 y_t^4$ and $y_t^6$ \cite{vanderBij:2000cg,Faisst:2003px}, which contribute to
the custodial symmetry breaking part of $\Delta r$. Finally, the 4-loop $g_3^6 y_t^2$ contribution to 
the $\rho$ parameter has been obtained in the on-shell scheme in refs.~\cite{Schroder:2005db,Chetyrkin:2006bj,Boughezal:2006xk}. (However, this 4-loop correction is small, and the remaining unknown 3-loop and 4-loop contributions to $G_F$ in the on-shell scheme are quite possibly larger.) The preceding effects have been summarized in a parameterized numerical fit
\cite{Awramik:2003rn} for the Standard Model prediction of $M_W$ resulting from the measured $G_F$, with a claimed theoretical uncertainty (not including parametric input uncertainties) on the order of 4 MeV.

Other schemes for organizing the radiative corrections in perturbation theory are useful and arguably provide advantages with respect to the on-shell scheme.
One alternative is the hybrid on-shell-\MSbar scheme proposed in
\cite{Degrassi:2014sxa}, which uses, instead of eq.~(\ref{eq:defineDeltar}),
\beq
G_F &=& \frac{g^2}{4 \sqrt{2} M_W^2} (1 + \Delta \hat r_W),
\label{eq:defineDeltarhatW}
\eeq
where $g$ is the running $SU(2)_L$ gauge coupling evaluated at the \MSbar renormalization scale $Q = M_Z$. This defines a quantity $\Delta \hat r_W$, which
was obtained to full 2-loop order in ref.~\cite{Degrassi:2014sxa} with the 3-loop leading QCD corrections. The resulting prediction for $M_W$ was found in \cite{Degrassi:2014sxa} to be lower than the on-shell determination by about 6 MeV, for fixed values of the other inputs. 

In refs.~\cite{Kniehl:2015nwa,Kniehl:2016enc} and in \cite{Martin:2019lqd}, two pure \MSbar scheme approaches are proposed, which differ in the way that they treat the Higgs vacuum expectation value (VEV). One motivation for using a pure \MSbar scheme is to facilitate a simple modular matching of on-shell observables onto the \MSbar parameters of the theory, which can in turn be matched to the Lagrangians for theories of new physics in the ultraviolet. In refs.~\cite{Kniehl:2015nwa,Kniehl:2016enc}, the Fermi constant is expressed as
\beq
G_F &=& \frac{1}{\sqrt{2} v^2_{\rm tree}} (1 + \Delta \overline{r}),
\label{eq:defineDeltarbar}
\eeq
which defines a quantity $\Delta \overline{r}$ in terms of the tree-level VEV
$v_{\rm tree}$ of the Higgs field in the \MSbar scheme. Writing the tree-level potential for the complex
Higgs field $H$ as
\beq
V = \Lambda + m^2 H^\dagger H + \lambda (H^\dagger H)^2,
\eeq
where $m^2<0$ and $\lambda$ is the Higgs self-coupling, one has
\beq
v_{\rm tree}^2 = -m^2/\lambda.
\eeq
Equation (\ref{eq:defineDeltarbar}) has the disadvantage that, since one is expanding around a VEV that does not minimize the full potential, one must include tadpole diagrams. The tadpole diagrams involve zero-momentum propagators of the Higgs boson, leading to factors of $\lambda$ in the denominators, so that the loop expansion parameter for diagrams involving the top quark is not the usual $N_c y^2_t/16 \pi^2$, but rather
$N_c y_t^4/16\pi^2 \lambda$,
leading to a slower convergence of perturbation theory due to the extra factor of $y_t^2/\lambda$ for each loop order. 

To avoid this, in ref.~\cite{Martin:2019lqd} and in the present paper one expands instead about the minimum of the VEV $v$ for the full effective potential in Landau gauge. In this tadpole-free pure \MSbar scheme, the sum of all tadpole diagram contributions (including the tree-level tadpoles) to any given observable vanishes identically,
and so there are no contributions in perturbation theory with the extra factors of $y_t^2/\lambda$. The use of $v$ rather than $v_{\rm tree}$ can be thought of as corresponding to a complete resummation of all tadpole diagrams; essentially, perturbation theory is improved by expanding about the correct vacuum. The effective potential
is known in Landau gauge at full 3-loop order in the Standard Model \cite{Ford:1992pn,Martin:2013gka,Martin:2017lqn}
(and in general theories \cite{Martin:2001vx,Martin:2017lqn}, including supersymmetric ones \cite{Martin:2023fno}), and the leading 4-loop QCD corrections are given in \cite{Martin:2015eia}. The two VEV definitions are related by 
\beq
v^2 = v_{\rm tree}^2 - \frac{1}{\lambda} \sum_{\ell=1}^\infty \frac{1}{(16 \pi^2)^\ell} \Delta_\ell,
\eeq
making explicit the $1/\lambda$ dependence,
where the $\Delta_\ell$ have been given in refs.~\cite{Martin:2017lqn,Martin:2015eia}.
The present paper is organized around
the following counterpart to eq.~(\ref{eq:defineDeltar}), (\ref{eq:defineDeltarhatW}), or (\ref{eq:defineDeltarbar}),
\beq
G_F &=& \frac{1}{\sqrt{2} v^2} (1 + \Delta \tilde r),
\label{eq:defineDeltartilde}
\eeq
which defines a quantity $\Delta \tilde r$. This directly relates the measured Fermi decay constant to the exact Landau gauge vacuum expectation value of the Higgs field. In this approach, the physical masses of the $W$ and $Z$ bosons are obtained as separate calculations, which have been carried out to full 2-loop order with 3-loop QCD corrections in refs.~\cite{Martin:2015lxa,Martin:2015rea,Martin:2022qiv}. This is part of a larger program, including also refs.~\cite{Martin:2014cxa,Martin:2016xsp,Martin:2018yow}, to relate the Standard Model observables to the \MSbar scheme Lagrangian parameters. All of these results are implemented consistently in the open-source computer code {\tt SMDR} \cite{Martin:2019lqd} and in a set of simple interpolating formulas in ref.~\cite{Alam:2022cdv}, which will be updated with the results of the present paper.

As a counterpoint to the obvious advantage of avoiding tadpole diagrams and their concomitant slower convergence of perturbation theory, the tadpole-free scheme has the apparent disadvantage that it is necessarily restricted to Landau gauge for the electroweak sector, because the effective potential is not known beyond 2-loop order \cite{Martin:2018emo} outside of Landau gauge. However, in a strict sense, this is not really a disadvantage, since physical quantities do not depend on the choice of gauge fixing. Moreover, the fact that one loses the opportunity to check calculations by observing the cancellation of gauge-fixing parameters is ameliorated by the fact that within Landau gauge there are similar nontrivial checks at 2-loop order and beyond, involving the cancellation, after resummation, of dependence on the unphysical Goldstone-boson squared mass including infrared singularities and potentially complex logarithms in individual diagrams.

The loop expansion for $\Delta \tilde r$ may be written as 
\beq
\Delta \tilde r &=& 
\frac{1}{16\pi^2} \Delta \tilde r^{(1)} +
\frac{1}{(16\pi^2)^2} \Delta \tilde r^{(2)} +
\frac{1}{(16\pi^2)^3} \Delta \tilde r^{(3)} + \cdots
.
\eeq
The complete expressions for $\Delta \tilde r^{(1)}$ and $\Delta \tilde r^{(2)}$ were given in ref.~\cite{Martin:2019lqd}. The purpose of the present paper is to provide the leading contributions to the 3-loop part $\Delta \tilde r^{(3)}$. Here, ``leading" means that in the 3-loop part I take 
\beq
g_3^2, y_t^2, \lambda &\gg& g^2, g^{\prime 2},
\label{eq:gaugeless}
\eeq
which is sometimes known as the ``gaugeless" limit. Although the effects of $\lambda$ are smaller than those of $g_3^2$ and $y_t^2$, it is natural to also treat it as large compared to the electroweak couplings, because $M_t^2, M_h^2 > M_W^2, M_Z^2$, and also because radiative corrections from lower loop orders involving the top-quark Yukawa coupling necessarily induce $\lambda$ anyway; for example, the 1-loop beta function for $\lambda$ contains a term proportional to 
$N_c y_t^4$. All other Yukawa couplings $y_b, y_\tau,\ldots$ are neglected in this paper (but are included
in the 1-loop correction $\Delta \tilde r^{(1)}$).

The limit (\ref{eq:gaugeless}) implies the important practical simplification that the contributions to $\Delta \tilde r^{(3)}$ all come only from the corrections to the $W$-boson self-energy at zero momentum, and do not involve any box diagrams or internal electroweak vector propagators except for the $W$-boson propagator at zero momentum in the case of one-particle reducible diagrams. Within this approximation, all effects involving the top-quark and Higgs-boson masses and couplings are treated exactly, and the final result for $\Delta \tilde r^{(3)}$ depends on only two squared mass scales,
which in the \MSbar scheme are 
\beq
t &=& y_t^2 v^2/2,
\\
h &=& 2 \lambda v^2.
\eeq 
The latter version of the running Higgs-boson squared mass differs slightly from the tree-level
Lagrangian expression,
\beq
H &=& m^2 + 3 \lambda v^2 = h + G,
\eeq
where the Goldstone-boson running squared mass at tree level in the tadpole-free scheme is
\beq
G = m^2 + \lambda v^2 .
\eeq
The dependences
on $G$ and $H$ (or, equivalently, $m^2$) are completely eliminated below in favor of $h$ by 
resummation \cite{Martin:2014bca,Elias-Miro:2014pca}, following the process exactly as described for the effective potential at 3-loop order in ref.~\cite{Martin:2017lqn}.

%%%%%%%%%%%%%%%%%%%%%%%%%%%%%%%%%%%%%%%%%%%%%%%%%%%%%%%%%%%%%%%
\section{Procedure\label{sec:procedure}}
\setcounter{equation}{0}
\setcounter{figure}{0}
\setcounter{table}{0} 
\setcounter{footnote}{1}

This section provides information about the procedure followed. As a first note,
I find it most efficient and pleasant to avoid the formalism of counterterm diagrams. Instead,
each Feynman diagram contributing to $(1 + \Delta \tilde r)/\sqrt{2} v^2$ up to 3-loop order is calculated in 
\beq
d = 4 - 2 \epsilon
\eeq
dimensions entirely in terms of bare couplings and masses. Then, renormalization is accomplished by simply rewriting the results from 0, 1, and 2 loops using the well-known relations between the bare parameters $v_B^2$, $g_{3B}$, $y_{tB}$, $\lambda_{B}$, and $m^2_B$ and the corresponding \MSbar parameters $v^2$, $g_3$, $y_t$, $\lambda$, and $m^2$, and then expanding to order $\epsilon^0$. 
(For the 3-loop diagrams, of course it does not
matter whether one uses the bare or \MSbar parameters, since the difference is of higher order.)
In general, the relation between each bare parameter $X_B$ and the corresponding renormalized parameter $X$
takes the form
\beq
X_B = \mu^{\rho_X \epsilon} \left (X + \sum_{\ell=1}^\infty \frac{1}{(16 \pi^2)^\ell} \sum_{n=1}^\ell \frac{c^X_{\ell,n}}{\epsilon^n} \right ) ,
\eeq
where the regularization scale $\mu$ is related to the \MSbar renormalization scale $Q$ by
\beq
Q^2 = 4\pi e^{-\gamma_E} \mu^2,
\eeq
where $\gamma_E$ is the Euler--Mascheroni constant, and  
\beq
\rho_{v^2} = -2,
\qquad 
\rho_{g_3} = \rho_{y_t} = 1,
\qquad 
\rho_\lambda = 2,
\qquad
\rho_{m^2} = 0 .
\eeq
The counterterm coefficients $c^X_{\ell,n}$
are polynomials in the \MSbar parameters. The ones necessary for the present paper in the gaugeless limit
of eq.~(\ref{eq:gaugeless}) are
\beq
c^{v^2}_{1,1} &=& -N_c y_t^2  v^2
,
\label{eq:cv211}
\\
c^{v^2}_{2,1} &=& \Bigl (-\frac{5}{2} N_c C_F g_3^2 y_t^2 + \frac{9}{8} N_c y_t^4 - 3 \lambda^2 
\Bigr ) v^2
,
\\ 
c^{v^2}_{2,2} &=& \Bigl (3 N_c C_F g_3^2 y_t^2 - \frac{3}{4} N_c y_t^4 
\Bigr ) v^2
,
\\ 
c^{v^2}_{3,1} &=& \biggl (
\Bigl [
\Bigl (6 \zeta_3 - \frac{77}{6} \Bigr )  C_G 
+ \Bigl (\frac{119}{12} - 12 \zeta_3 \Bigr ) C_F 
+ \frac{16}{3} n_g T_F
\Bigr ] N_c C_F g_3^4 y_t^2 
+ \Bigl [6 \zeta_3 - \frac{5}{8} \Bigr ] N_c C_F g_3^2 y_t^4 
\nonumber \\ && 
+ \Bigl [\frac{25}{48} - \zeta_3 - 2 N_c \Bigr ] N_c y_t^6 
- 5 N_c \lambda y_t^4
+ \frac{15}{2} N_c \lambda^2 y_t^2
+ 12 \lambda^3 
\biggr ) v^2
,
\\ 
c^{v^2}_{3,2} &=& 
\biggl (
\Bigl [
\frac{83}{6}  C_G 
+ 6 C_F 
- \frac{20}{3} n_g T_F
\Bigr ] N_c C_F g_3^4 y_t^2 
- \frac{39}{4} N_c C_F g_3^2 y_t^4 
\nonumber \\ && 
+ \Bigl [ \frac{5}{8} + \frac{9}{8} N_c \Bigr ] N_c y_t^6 
+6 N_c \lambda y_t^4
-3 N_c \lambda^2 y_t^2
-24 \lambda^3 
\biggr ) v^2
,
\\ 
c^{v^2}_{3,3} &=& 
\biggl (
\Bigl [
-\frac{11}{3}  C_G 
- 6 C_F 
+ \frac{8}{3} n_g T_F
\Bigr ] N_c C_F g_3^4 y_t^2 
+ \frac{9}{2} N_c C_F g_3^2 y_t^4 
%\nonumber \\ && 
- \Bigl [\frac{3}{4} + \frac{1}{4} N_c \Bigr ] N_c y_t^6 
\biggr ) v^2
,
\eeq
and
\beq
c^{y_t}_{1,1} &=& -3 C_F g_3^2 y_t + \left (\frac{3}{4} +  \frac{1}{2} N_c \right ) y_t^3 
,
\\
c^{y_t}_{2,1} &=& \left (-\frac{97}{12} C_G - \frac{3}{4} C_F  + \frac{10}{3} n_g T_F \right ) C_F g_3^4 y_t
+ \left (3 + \frac{5}{4} N_c \right )C_F g_3^2 y_t^3
\nonumber \\ &&
+ \left (\frac{3}{8} - \frac{9}{8} N_c \right ) y_t^5
- 3 \lambda y_t^3 + \frac{3}{2} \lambda^2 y_t
,
\\
c^{y_t}_{2,2} &=& \left (\frac{11}{2} C_G + \frac{9}{2} C_F  - 4 n_g T_F \right ) C_F g_3^4 y_t
- \left (\frac{9}{2} + 3 N_c \right ) C_F g_3^2 y_t^3
\nonumber \\ &&
+ \left (\frac{27}{32} + \frac{9}{8} N_c  + \frac{3}{8} N_c^2 \right ) y_t^5
,
\eeq
and
\beq
c^{\lambda}_{1,1} &=& -N_c y_t^4 + 2 N_c y_t^2 \lambda + 12 \lambda^2,
\\
c^{\lambda}_{2,1} &=& -2 N_c C_F g_3^2 y_t^4 + 5 N_c C_F g_3^2 y_t^2 \lambda
+ \frac{5}{2} N_c y_t^6 - \frac{1}{4} N_c y_t^4 \lambda - 12 N_c y_t^2\lambda^2
- 78 \lambda^3, 
\\
c^{\lambda}_{2,2} &=& 6 N_c C_F g_3^2 y_t^4 -6 N_c C_F g_3^2 y_t^2 \lambda
- \left (\frac{3}{2} + 2 N_c \right ) N_c y_t^6 + \left (3 N_c- \frac{21}{2} \right ) N_c y_t^4 \lambda \nonumber \\ &&
+ 36 N_c y_t^2\lambda^2
+144 \lambda^3, 
\eeq
and
\beq
c^{m^2}_{1,1} &=& \left (N_c y_t^2 + 6 \lambda \right ) m^2
,
\\
c^{m^2}_{2,1} &=& \left (\frac{5}{2} N_c C_F g_3^2 y_t^2 - \frac{9}{8} N_c y_t^4 - 6 N_c y_t^2 \lambda 
- 15 \lambda^2  \right ) m^2
,
\\
c^{m^2}_{2,2} &=& \left (-3 N_c C_F g_3^2 y_t^2 + \Bigl [N_c - \frac{9}{4} \Bigr ] N_c y_t^4 
+ 12 N_c y_t^2 \lambda
+54 \lambda^2  \right ) m^2
\eeq
and
\beq
c^{g_3}_{1,1} &=& \left ( -\frac{11}{6} C_G + \frac{4}{3} n_g T_F \right ) g_3^3
.
\label{eq:cg311}
\eeq
In the Standard Model one has $C_G = N_c = 3$ is the number of colors, and $C_F = 4/3$ and $T_F = 1/2$ are the quadratic Casimir and Dynkin index of the fundamental representation of $SU(3)_c$, and $n_g = 3$ is the number of
generations.

\begin{figure}[p]
\includegraphics[width=0.95\linewidth]{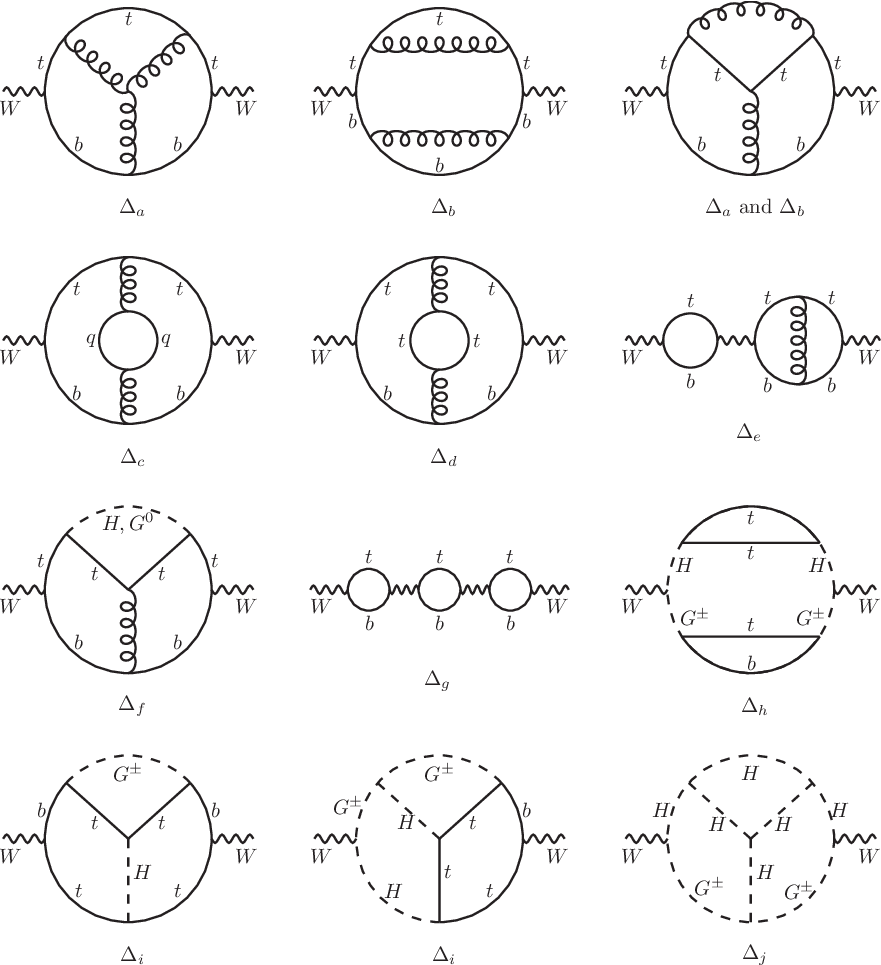}
\caption{\label{fig:diagramexamples} Some sample 3-loop diagrams contributing, as labeled, to
$\Delta_a$, $\Delta_b$, \ldots, $\Delta_j$ in eq.~(\ref{eq:Deltartilde3split}).}
\end{figure}

In order to facilitate maximally informative checks, the final result for the 3-loop contribution $\Delta\tilde r^{(3)} $ is split into 10 distinct parts $\Delta_a,\Delta_b,\ldots,\Delta_j$ with different QCD $SU(3)$ group-theory invariants and numbers of fermion loops, as
\beq
\Delta \tilde r^{(3)} &=& 
g_3^4 y_t^2 N_c C_F \left (C_G \Delta_a + C_F \Delta_b + n_g T_F \Delta_c + T_F \Delta_d \right )
\nonumber \\ &&
+ g_3^2 y_t^4 N_c C_F (N_c \Delta_e + \Delta_f)
+ y_t^6 N_c (N_c^2 \Delta_g + N_c \Delta_h + \Delta_i)
+ \lambda^3 \Delta_j
.
\label{eq:Deltartilde3split}
\eeq
All diagrams contributing are vacuum integrals (because of the vanishing of the external momentum) with up to 3 loops occurring in the self-energy of the $W$ boson.
For illustration, some sample 3-loop diagrams contributing to each of these parts are shown in Figure \ref{fig:diagramexamples}. In one-particle-reducible diagrams like the ones shown for $\Delta_e$ and $\Delta_g$, the extra factors of $g^2$ from the internal $W$ boson couplings are canceled by the corresponding zero-momentum $W$ propagators proportional to $4/g^2 v^2$. The diagram shown for $\Delta_g$ is the unique 3-loop one contributing proportional to $N_c^3$ due to three fermion loops. Note that the part $\Delta_i$ has a factor of $y_t^6$ extracted, but the contributing 3-loop diagrams all involve virtual Higgs and Goldstone bosons, and as illustrated by the two example diagrams shown, some parts of these contributions could instead more naturally be associated with coupling factors $y_t^4 \lambda$ or $y_t^2 \lambda^2$. However, this separation is not made, as there is not always a unique way to assign contributions to these different coupling factors; a factor of $h/t$ in a term arising from the integration can always be traded for $4 \lambda/y_t^2$ in the couplings multiplying it.
The contribution $\Delta_j$, proportional to $\lambda^3$, is due to all diagrams involving no fermion loops (only virtual scalar bosons), of which only one example is shown. 

%%%%%%%%%%%%%%%%%%%%%%%%%%%%%%%%%%%%%%%%%%%%%%%%%%%%%%%%%%%%%%%%%%%%%%%%%%%%%%%%%
%%%%%%%%%%%%%%%%%%%%%%%%%%%%%%%%%%%%%%%%%%%%%%%%%%%%%%%%%%%%%%%%%%%%%%%%%%%%%%%%%
%%%%%%%%%%%%%%%%%%%%%%%%%%%%%%%%%%%%%%%%%%%%%%%%%%%%%%%%%%%%%%%%%%%%%%%%%%%%%%%%%
\begin{figure}[p]
\includegraphics[width=0.95\linewidth]{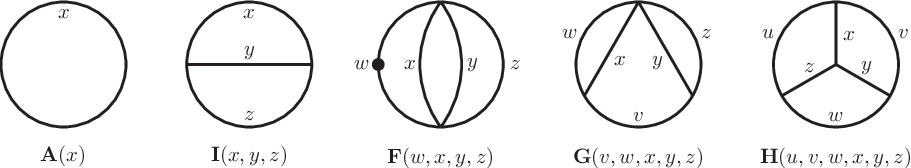}
\caption{\label{fig:basistopologies} Topologies for the 1-loop, 2-loop, and 3-loop master vacuum scalar integrals with squared masses $u,v,w,x,y,z$. The dot on the ${\bf F}(w,x,y,z)$ diagram denotes a doubled propagator. The precise definitions are given in ref.~\cite{Martin:2016bgz}, which also provides a computer code {\tt 3VIL} for their numerical evaluation.}
\end{figure}
%%%%%%%%%%%%%%%%%%%%%%%%%%%%%%%%%%%%%%%%%%%%%%%%%%%%%%%%%%%%%%%%%%%%%%%%%%%%%%%%%

After computing the color and fermion index traces and Lorentz vector index algebra, the resulting bare integrals are reduced, using the program {\tt KIRA} \cite{Maierhofer:2017gsa,Maierhofer:2018gpa,Klappert:2020nbg}, to a set of master integrals in $d$
dimensions. The master integrals are chosen from among the ones defined, using the notations and conventions of ref.~\cite{Martin:2016bgz}, as ${\bf A}(x)$,
${\bf I}(x,y,z)$, ${\bf F}(w,x,y,z)$, ${\bf G}(v,w,x,y,z)$, and ${\bf H}(u,v,w,x,y,z)$. These are pictured
in Figure \ref{fig:basistopologies}, and their precise definitions are given in ref.~\cite{Martin:2016bgz}. 
In the present calculation, each of the arguments $u,v,w,x,y,z$ is chosen from among the squared masses 0, $t$, $H$, $G$ (or, for 1- and 2-loop diagrams, the corresponding bare squared masses). Since these squared masses are not generic, there are numerous non-trivial relations between these integrals, allowing some of them to be eliminated. A list of pertinent identities between
the candidate master integrals, also obtained or checked using {\tt KIRA}, is provided in the ancillary file {\tt identities.txt} provided with this paper. 

When using these identities it is important to choose the uneliminated master integrals in such a way that factors of $d-4$ do not appear in denominators, to avoid unwanted and unnecessary divergences when expanding in small $\epsilon$. This means that the master integrals are chosen in such a way as to form an $\epsilon$-finite basis, in the sense of ref.~\cite{Chetyrkin:2006dh}.
As one typical example, consider the identity
\beq
{\bf G}(0,0,0,0,x) &=& \Bigl [
(d-4) x {\bf H} (0,0,0,0,x,x) 
+ \frac{4}{3} (4 - d) x  {\bf H}(0,0,x,0,x,x)
\nonumber \\ && 
+ \frac{(d-2)^2}{6 (3-d) x^2} [{\bf A}(x)]^3 
+ \frac{2-d}{x} {\bf A}(x) {\bf I}(0,0,x) \Bigr ]/(3d-10) .
\eeq
One can use this to eliminate ${\bf G}(0,0,0,0,x)$ in favor of the others, but one should not use
it to instead eliminate ${\bf H} (0,0,0,0,x,x)$ or ${\bf H}(0,0,x,0,x,x)$, as this would introduce
unnecessary additional poles in $\epsilon$.

Another slightly tricky feature of the master integral choice is that one should strive to eliminate ${\bf F}$ integrals for which the first squared mass argument vanishes, as these contain infrared divergences associated with the doubled
massless propagator, making the $\epsilon \rightarrow 0$ limit more difficult. In the present calculation, two types of such integrals occur. The easiest are 
those in which one of the other propagator masses also vanishes. These can be trivially eliminated using
the remarkably simple identity
\beq
{\bf F}(0,0,x,y) &=& (3-d) {\bf G}(0,0,0,x,y) .
\eeq
The second type is ${\bf F}(0,x,x,y)$, which occurs in $\Delta_f$. Eliminating this is more difficult, but one way to do it uses
the identity
\beq
{\bf F}(0,x,x,y) &=& \frac{d-3}{(8d-24) x^2 - 4 x y + (d-2) y^2} 
\Bigl [
(4 x^2 - 2 x y + y^2) \Bigl \{
(d-2) {\bf A}(x) {\bf I}(x,x,y)/x
\nonumber \\ &&
+ (3d/2 - 5) {\bf G}(0,x,x,x,x) 
+ (d-4)(2x - 3 y/2) {\bf H}(0,x,x,x,y,x) 
\nonumber \\ &&
+ (y - 4 x) y \frac{\partial }{\partial y} {\bf H}(0,x,x,x,y,x)  
\Bigr \}
+ (2 - d) (2 x - y) [{\bf A}(x) - {\bf A}(y)] {\bf I} (0, 0, x)
\nonumber \\ &&
+ (2 - d) (x - y) (2 x - y) \left [ {\bf G}(0, 0, x, x, y) + {\bf A}(x) {\bf I}(0,x,y)/x \right ]
\nonumber \\ &&
+ [(d-2)^2/(d-3)] {\bf A}(x)^2 {\bf A}(y)
\Bigr ] 
.
\label{eq:F0xxy}
\eeq
Note that this implies taking one of the uneliminated bare master integrals to be 
$\frac{\partial }{\partial y} {\bf H}(0,x,x,x,y,x)$. This special case is at variance with the otherwise universal strategy of only choosing master integrals from among those of types 
${\bf A}$, ${\bf I}$, ${\bf F}$, 
${\bf G}$, and ${\bf H}$, and not their derivatives. Fortunately, expanding in $\epsilon$, the integral
$\frac{\partial }{\partial y} {\bf H}(0,x,x,x,y,x)$ has no poles in $\epsilon$, and its $\epsilon^0$ part 
has a known expression in terms of other renormalized master integrals, to be given very shortly.

After the reduction to master integrals just described, but before proceeding further, it is a useful check that the dependence on the QCD gauge-fixing parameter vanishes separately for each of the contributions $\Delta_a,\Delta_b,\Delta_c,\Delta_d,\Delta_e$, and $\Delta_f$.

The next step is to expand both the bare couplings and the bare master integrals in $\epsilon$, keeping terms in $G_F$ up to $\epsilon^0$. This makes use of the results of section II of ref.~\cite{Martin:2016bgz}, which writes the bare master integrals
in terms of renormalized master integrals $A(x)$, $I(x,y,z)$, $F(w,x,y,z)$, $G(v,w,x,y,z)$ and $H(u,v,w,x,y,z)$, which have no dependence on $\epsilon$. Note that the renormalized master integrals have the same names as the corresponding
bare master integrals, but with bold-faced letters replaced by non-bold-faced letters. 
(However, the renormalized master integrals $I$, $F$ and $G$ are not simply the finite parts of the bare integrals ${\bf I}$, ${\bf F}$, and ${\bf G}$ as $\epsilon \rightarrow 0$.)
The renormalized
master integrals also depend logarithmically on the \MSbar renormalization scale $Q$, which is not listed explicitly among the arguments because it is the same everywhere in a given calculation.
As explained in ref.~\cite{Martin:2016bgz}, it is also convenient to introduce an infrared-safe renormalized integral 
\beq
\overline{F}(0,x,y,z) = \lim_{w \rightarrow 0} \left [F(w,x,y,z) + \lnbar(w) I(x,y,z) \right ],
\eeq
where 
\beq
\lnbar(x) &=& \ln(x/Q^2) \,=\, A(x)/x + 1.
\eeq
At this point one can deal with the special case of eq.~(\ref{eq:F0xxy}) by using the identity
\beq
&&
y\frac{\partial }{\partial y} H(0,x,x,x,y,x) \>=\>
\frac{1}{(4x-y)^2} \Bigl [
4(x-y) [G(x,0,x,x,y) + F(y,0,x,x)]
\nonumber \\ &&\qquad\qquad 
+ (3y - 4 x) G(y,x,x,x,x) 
+ (2y - 8x) \overline{F}(0,x,x,y)
+ 7 (y - 4 x) F(x,0,0,x)  
\nonumber \\ &&\qquad\qquad 
+ (4 A(x) + 4x - 2 y) I(x,x,y)
+ 3 (4 x - y) A(x)^3/x^2
+ (6/x - 4/y) A(x)^2 A(y)
\nonumber \\ &&\qquad\qquad 
+ (2y/x - 12) A(x)^2
+ (8x/y - 12) A(x) A(y)
+ (17y/4 - 21 x) A(x)
\nonumber \\ &&\qquad\qquad 
+ (3x + 4 y - 8 x^2/y) A(y)
+ 48 x^2 + 28 x y/3 - 26 y^2/3
\Bigr ] 
%\nonumber \\ && \qquad\qquad
%+ {\cal O}(\epsilon)
,
\eeq
which can be obtained by first computing $\frac{\partial }{\partial y} H(0,x,x,x,y,z)$, and then taking the limit $z \rightarrow x$, using the results in the ancillary file {\tt derivatives.txt} of ref.~\cite{Martin:2016bgz}.
It is also convenient 
to replace the renormalized master integrals that only involve one squared mass scale $x$ by their analytic
expressions given in section V of ref.~\cite{Martin:2016bgz}, including some relevant results found previously in 
refs.~\cite{Broadhurst:1991fi,Avdeev:1995eu,Broadhurst:1998rz,Fleischer:1999mp,Steinhauser:2000ry,
Schroder:2005va}, which involve known numerical constants and
$\lnbar(x)$.
The irrational constants [besides $\pi$ and $\ln(2)$ and $\sqrt{3}$] appearing in the one-scale renormalized master integrals pertinent for the present paper are
\beq
\zeta_3 &=& 1.2020569031595943\ldots,
\\
{\rm Li}_4(1/2) &=& 0.5174790616738994\ldots,
\\
{\rm Ls}_2 &=&
0.6766277376064358\ldots.
\eeq
After combining the 3-loop contributions with those from the expansion of bare parameters in lower loop orders, the cancellation
of all poles proportional to $1/\epsilon^n$ for $n = 1,2,3$ for each of the 10 separate contributions
$\Delta_a,\Delta_b,\ldots,\Delta_j$ provides another powerful check. In particular, this is a check that the present
calculation is consistent with the well-known Standard Model counterterms for the \MSbar quantities.

The final step is the expansion and resummation \cite{Martin:2014bca,Elias-Miro:2014pca}
in the Goldstone-boson squared mass, in which $G$ is eliminated  (along with the elimination of $H = h+G$ in favor of $h$). As noted above,
this was done following the process described in section V.B of ref.~\cite{Martin:2017lqn}. This provides more independent checks, as some of the individual diagrams have $G \rightarrow 0$ singularities proportional to $1/G$ and $\lnbar(G)$ and $\lnbar^2(G)$, which successfully cancel in the complete result. As remarked above, when fixed to Landau gauge this is the counterpart of the check that would follow from cancellation of the gauge-fixing parameter $\xi$ if the calculation were performed in an $R_\xi$ or similar gauge. 

The final results, given in the next section, are expressed in terms of $\lnbar(t)$ and $\lnbar(h)$ and renormalized $\epsilon$-finite master integrals $I(x,y,z)$,
$F(w,x,y,z)$, $\overline{F}(0,x,y,z)$, $G(v,w,x,y,z)$, and $H(u,v,w,x,y,z)$, where each squared mass
argument is 0, $t$, or $h$. Another practical check is that in this final result there is a complete cancellation of the functions appearing beyond order $\epsilon^0$ in the expansions of the bare integrals ${\bf A}(x)$ and ${\bf I}(x,y,z)$. This agrees with the general argument in ref.~\cite{Martin:2021pnd} from the fact that the bare master integrals were chosen as an $\epsilon$-finite \cite{Chetyrkin:2006dh} basis. 

In the expression for $\Delta_i$, some of the individual coefficients of the master integrals have denominators proportional to 
$h-4t$, $h - 3 t$, $h - 2 t$, and $h - t$. As yet another consistency check, I find that the total result is nevertheless finite in the (not phenomenologically realistic) limits that each of these quantities vanishes, due to non-trivial relations between the different master integrals at those special threshold values for $h$, which were verified numerically using {\tt 3VIL} \cite{Martin:2016bgz}. The same holds for terms in $\Delta_f$ with denominators proportional to $h-4t$ and $h-t$.

%%%%%%%%%%%%%%%%%%%%%%%%%%%%%%%%%%%%%%%%%%%%%%%%%%%%%%%%%%%%%%%
\section{Results\label{sec:results}}
\setcounter{equation}{0}
\setcounter{figure}{0}
\setcounter{table}{0} 
\setcounter{footnote}{1}

In the gaugeless limit approximation of eq.~(\ref{eq:gaugeless}), the 1-loop and 2-loop contributions
to $\Delta \tilde r$ are
\beq
\Delta \tilde r^{(1)} &=& N_c y_t^2 \left [\lnbar(t) - 1/2 \right ] - \lambda
,
\label{eq:Deltar1gaugeless}
\\
\Delta \tilde r^{(2)} &=& N_c C_F g_3^2 y_t^2 \left [\frac{\pi^2}{3} - \frac{13}{4} + 2 \lnbar(t) - 3 \lnbar^2(t) \right ]
+ N_c^2 y_t^4 \bigr [ \lnbar(t) - 1/2 \bigr ]^2
\nonumber \\ &&
+ N_c y_t^4 \biggl [ 
\left (\frac{\pi^2}{4} + \frac{15}{4} \right ) \frac{t}{h} + \frac{3}{16} - \frac{\pi^2}{24} - \frac{2h}{t}
+ \left (\frac{h}{4t}-\frac{3t}{h} - \frac{7}{2} \right ) \lnbar(t)
\nonumber \\ &&
+ \left (\frac{2h}{t} - \frac{3}{2} \right ) \lnbar(h)
+ \left (\frac{3}{8} + \frac{3t}{4h} \right ) \lnbar^2(t)
+ \left (\frac{3}{2} - \frac{3h}{4t} \right ) \lnbar(t) \lnbar(h)
\nonumber \\ &&
+ \left (\frac{3}{2h} - \frac{9}{4t} + \frac{3h}{4 t^2} \right ) I(0,h,t)
+ \left (\frac{3}{2t} - \frac{3 h}{4 t^2} \right ) I(h,t,t)
\biggr ]
\nonumber \\ &&
+ \lambda^2 \left [\frac{49}{2} - \frac{\pi^2}{3} - 18 \sqrt{3} {\rm Ls}_2 - 6 \lnbar(h) \right ]
.
\label{eq:Deltar2gaugeless}
\eeq
In practical applications one uses the more complete expressions including the effects of $g, g'$ at 2-loop order (and even the $y_b$ and $y_\tau$ contributions in the 1-loop part). Those were given in ref.~\cite{Martin:2019lqd}.

I now present the results for $\Delta \tilde r^{(3)}$ in the form specified by eq.~(\ref{eq:Deltartilde3split}). Six of the ten contributions depend on only the single mass scale $t = y_t^2 v^2/2$, which appears only logarithmically because $ \Delta \tilde r$ is dimensionless. First,
the contributions at leading order in QCD, proportional to $g_3^4 y_t^2$, are
\beq
\Delta_a &=& 
-\frac{869}{24} + \frac{73 \pi^2}{18} + 100 \zeta_3 - \frac{391 \pi^4}{360} 
+ \frac{16}{3} \left [\ln^2(2) - \pi^2 \right ] \ln^2(2) 
- \frac{117}{2} \sqrt{3} {\rm Ls}_2
- \frac{27}{2} ({\rm Ls}_2)^2 
\nonumber \\ && 
+ 128 {\rm Li}_4(1/2) 
+ \left (\frac{137}{4} - \frac{11 \pi^2}{9} - 18 \zeta_3\right) \overline\ln(t) 
- \frac{119}{6} \overline\ln^2(t)
+ \frac{11}{3} \overline\ln^3(t)
,
\\
\Delta_b &=& 
\frac{1025}{36} + \frac{259 \pi^2}{54} -\frac{794}{3} \zeta_3 + \frac{109 \pi^4}{60} 
- \frac{32}{3} \left [\ln^2(2) - \pi^2 \right ] \ln^2(2) 
+ 117 \sqrt{3} {\rm Ls}_2
\nonumber \\ && 
+ 27 ({\rm Ls}_2)^2 
-256 {\rm Li}_4(1/2) 
+ \left (-\frac{95}{4} - 2 \pi^2 + 36 \zeta_3\right) \overline\ln(t) 
+ \frac{21}{2} \overline\ln^2(t)
+ 6 \overline\ln^3(t)
,
\\
\Delta_c &=& \frac{73}{3} -\frac{28\pi^2}{9} + 32 \zeta_3 + \left (-18 + \frac{8 \pi^2}{9} \right ) \overline\ln(t) + \frac{28}{3} \overline\ln^2(t)
- \frac{8}{3} \overline\ln^3(t)
,
\\
\Delta_d &=& 
-\frac{176}{3} + \frac{14\pi^2}{9} + \frac{368}{3} \zeta_3 - 108 \sqrt{3} {\rm Ls}_2 
.
\eeq
The contribution from diagrams containing a gluon and two separate fermion loops is
\beq
\Delta_e &=& \frac{13}{4} - \frac{\pi^2}{3} + \left (-\frac{17}{2} + \frac{2 \pi^2}{3} \right ) \overline\ln(t) 
+ 7 \overline\ln^2(t)
- 6 \overline\ln^3(t) ,
\eeq
and the contribution from the one diagram with three separate top/bottom loops is
\beq
\Delta_g &=& \left [\overline\ln(t) - 1/2 \right ]^3 .
%-\frac{1}{8} + \frac{3}{4} \overline\ln(t) 
%-\frac{3}{2} \overline\ln^2(t)
%+ \overline\ln^3(t) 
\eeq
[This generalizes to the contribution from $\ell$-loop order with the maximal number of fermion loops, namely $\ell$ separate 1-loop top/bottom subdiagrams,
which due to the chain structure is simple to evaluate as 
\beq
\Delta \tilde r^{(\ell)} \Bigl |_{\mbox{non-QCD leading $N_c$}} &=&
\bigl [ N_c y_t^2 (\lnbar(t) - 1/2) \bigr ]^\ell ,
\eeq
as in the $N_c y_t^2$ part of eq.~(\ref{eq:Deltar1gaugeless}) and the $N_c^2 y_t^4$ part of eq.~(\ref{eq:Deltar2gaugeless}).
However, this is of limited utility, as it gives a negligible contribution to $\Delta \tilde r$ for $\ell\geq 4$.]

Another relatively simple result is the one from all diagrams with only Higgs and Goldstone internal loop propagators (no fermion loops). After Goldstone boson squared mass resummation, this contribution depends only logarithmically on the single squared mass $h = 2 \lambda v^2$. It is
\beq
\Delta_j &=& 
-\frac{2809}{2} +\frac{517}{9} \pi^2 - \frac{712}{3} \zeta_3 + \frac{617 \pi^4}{45} 
- \frac{16\pi^3}{3\sqrt{3}}
- 64 \left [\ln^2(2) - \pi^2 \right ] \ln^2(2) 
+414 \sqrt{3} {\rm Ls}_2
\nonumber \\ && 
- 108 ({\rm Ls}_2)^2 
-1536 {\rm Li}_4(1/2) 
+ \left (630 - 8 \pi^2 - 432 \sqrt{3} {\rm Ls}_2 \right) \overline\ln(h) 
- 72 \overline\ln^2(h)
.
\eeq

That leaves the three contributions $\Delta_f$, $\Delta_h$, and $\Delta_i$, which each depend on two-scale renormalized $\epsilon$-finite master integrals.
Specifically, $\Delta_f$ contains the integrals
\beq
&&
I(0, h, t),\quad\!\! 
I(h, t, t),\quad\!\!
F(h, 0, 0, t),\quad\!\! 
F(h, 0, t, t),\quad\!\! 
\overline{F}(0, 0, h, t),\quad\!\! 
\overline{F}(0, h, t, t),\quad\!\! 
G(0, 0, t, h, t),\quad\!\!
\nonumber \\ && 
G(h, 0, t, t, t),\quad\!\! 
G(t, 0, 0, h, t),\quad\!\! 
H(0, 0, h, 0, t, t),\quad\!\! 
H(0, 0, t, t, h, t),\quad\!\! 
H(0, t, t, t, h, t)
,
\label{eq:neededf}
\eeq
and $\Delta_h$ depends on the integrals
\beq
I(0, h, t),\quad\!\! 
I(h, t, t),\quad\!\!
G(h, 0, t, 0, t),\quad\!\! 
G(h, 0, t, t, t),\quad\!\! 
G(h, t, t, t, t),\quad\!\! 
\eeq
while $\Delta_i$ contains the integrals
\beq
&&
I(0, h, t),\quad\!\! 
I(h, h, t),\quad\!\! 
I(h, t, t),\quad\!\!
F(h, 0, 0, t),\quad\!\! 
F(h, 0, t, t),\quad\!\! 
F(h, h, t, t),\quad\!\! 
F(t, 0, h, h),\quad\!\! 
\nonumber \\ && 
\overline{F}(0, 0, h, t),
\quad G(h, 0, 0, 0, t),\quad\!\! 
G(h, 0, 0, t, t),\quad\!\! 
G(h, 0, t, h, h),\quad\!\! 
G(h, 0, t, t, t),\quad\!\!
\nonumber \\ &&  
G(h, h, h, t, t),\quad\!\! 
G(h, t, t, t, t),\quad\!\! 
G(t, 0, 0, 0, h),\quad\!\! 
G(t, 0, 0, h, t),\quad\!\! 
G(t, 0, h, h, t),\quad\!\! 
\nonumber \\ && 
G(t, h, t, h, t),\quad\!\! 
H(0, 0, 0, 0, h, t),\quad\!\! 
H(0, 0, 0, h, t, t),\quad\!\! 
H(0, 0, h, t, 0, t),\quad\!\! 
H(0, 0, h, t, h, t),\quad\!\! 
\nonumber \\ && 
H(0, 0, t, h, t, t),\quad\!\! 
H(0, 0, t, t, h, t),\quad\!\! 
H(0, h, h, t, h, t),\quad\!\! 
H(0, h, t, t, t, h),\quad\!\! 
H(h, h, t, h, t, t),\quad\!\! 
\nonumber \\ && 
H(h, t, t, t, h, t). 
\label{eq:neededi}
\eeq
It should be noted that the choices of which renormalized integrals appear in these expressions are not unique, because of the identities that relate them due to the non-generic squared mass arguments.
The explicit expressions for $\Delta_f$, $\Delta_h$, and $\Delta_i$ are somewhat unwieldy and of little use to the human eye, and so they are relegated to the ancillary file {\tt Deltartilde3.txt} distributed with this paper, which for convenience also includes the other contributions $\Delta_a, \Delta_b, \ldots,\Delta_j$. 

Besides the consistency checks mentioned in the previous section, I have also checked that the results  satisfy the conditions following from the renormalization group invariance of $G_F$,
\beq
Q \frac{\partial}{\partial Q} \Delta \tilde r^{(\ell)}
&=& 
\beta_{v^2}^{(\ell)}/v^2 + 
\sum_{n = 1}^{\ell - 1} 
\left [ 
\beta_{v^2}^{(n)}/v^2
- \sum_X \beta_X^{(n)} \frac{\partial}{\partial X}
\right ] \Delta \tilde r^{(\ell-n)},
\label{eq:betacheck}
\eeq
for each of $\ell = 1,2,3$,
where $X$ is summed over $v^2$, $g_3$, $y_t$, $\lambda$ for $\ell=3$, and also $g, g'$ for $\ell=1,2$.
Here, the beta function for each parameter $X$ is
\beq
Q \frac{dX}{dQ} &=& \sum_{\ell=1}^\infty \frac{1}{(16\pi^2)^\ell} \beta_X^{(\ell)}
,
\eeq 
and can be found in the gaugeless limit from eqs.~(\ref{eq:cv211})-(\ref{eq:cg311}), by using $\beta_X^{(\ell)} = 2 \ell c^X_{\ell,1}$. 
In principle, eq.~(\ref{eq:betacheck}) is redundant with the earlier check of the cancellation of $1/\epsilon$ poles, but in practice it
is a useful check on the intermediate steps leading to the final result. 

The required two-scale integrals in eqs.~(\ref{eq:neededf})-(\ref{eq:neededi}) can be evaluated numerically on demand by the computer code {\tt 3VIL} \cite{Martin:2016bgz}.
In the current version 2.02 of {\tt 3VIL}, almost all of them are computed either in terms of polylogarithms or in terms of convergent series, including formulas found earlier in
refs.~\cite{Davydychev:2003mv,Kalmykov:2005hb,Bekavac:2009gz,Grigo:2012ji}. The only exceptions are that 
one Runge-Kutta run is necessary for $H(h,t,t,t,h,t)$, $G(t,h,t,h,t)$, $G(h,t,t,t,t)$, and $F(h,h,t,t)$,
and a second Runge-Kutta run is required for $H(h,h,t,h,t,t)$, and $G(h,h,h,t,t)$.
The total numerical evaluation time for the complete set of integrals is well under 1 second on modern computer hardware.
The new result for $\Delta \tilde r$ has been incorporated in the default evaluation of $G_F$ in version 1.3 of {\tt SMDR} \cite{Martin:2019lqd}.

To illustrate the impact of the 3-loop corrections found in this paper, consider a set of benchmark Standard Model on-shell parameters given by the latest (as of July 2025) central values from the Particle Data Group,
\beq
&&
G_F \>=\> 1.1663782 \times 10^{-5}\>{\rm GeV}^{-2},
\label{eq:GFbenchmark}
\\
&&
\alpha \>=\> 1/137.035999178,
\qquad
\Delta\alpha_{\rm had}^{(5)}(M_Z) \>=\> 0.02783,
\qquad
\alpha_S^{(5)}(M_Z) = 0.1180,
\\
&&
M_t \>=\> 172.4\>{\rm GeV},
\qquad 
M_h \>=\> 125.20\>{\rm GeV},
\qquad 
M_{Z} \>=\> 91.1876\>{\rm GeV},
\\
&&
m_b(m_b) \>=\> 4.183\>{\rm GeV}, 
\qquad
m_c(m_c) \>=\> 1.273\>{\rm GeV},
\qquad
m_s(2\>{\rm GeV}) \>=\> 0.0935\>{\rm GeV},
\phantom{xxxx}
\\
&&
m_d(2\>{\rm GeV}) \>=\> 0.00470\>{\rm GeV},
\qquad
m_u(2\>{\rm GeV}) \>=\> 0.00216\>{\rm GeV},
\qquad
\\
&&
M_{\tau} \>=\> 1.77693\>{\rm GeV},
\qquad
M_\mu \>=\> 0.1056583755 \>{\rm GeV},
\qquad
M_e = 0.00051099895 \>{\rm GeV}.
\label{eq:leptonsbenchmark}
\eeq
Using the utility {\tt calc\_fit} of {\tt SMDR} 1.3 with default settings, which includes the results of the present paper, the corresponding \MSbar parameters at $Q = 172.4$ GeV are found to be
\beq
%\mbox{MSbar couplings go here.}
Q &=& 172.4\>{\rm GeV},
\label{eq:referencemodelMSbarstart}
\\
g_3 &=& 1.16350130213,\quad\>\>
  g \>=\> 0.64759346159,\quad\>\>\,
 g' \>=\> 0.35859479822,
\\
    v &=&  246.600295864,\quad\>\>
\lambda  \>=\> 0.12629692329,\quad\>\>
m^2 \>=\> -(92.8716879304\>{\rm GeV})^2,
\label{eq:referencemodelMSbarhiggs}
\\
y_t    &=& 0.93097154656,\quad\>\>
y_b    \>=\> 0.01550800810,\quad\>\>
y_c    \>=\> 0.00340027571,
\\
y_s    &=& 0.00029291582,\quad\>\>
y_d    \>=\> 1.4724091371 \times 10^{-5},\quad\>\>
y_u    \>=\> 6.7321435654 \times 10^{-6},
\\
y_\tau &=& 0.00999493237,\quad\>\>
y_\mu  \>=\> 5.8838465029 \times 10^{-4},\quad\>\>
y_e    \>=\> 2.7929943407 \times 10^{-6}.
\phantom{xxxx}
\label{eq:referencemodelMSbarend}
\eeq
(Many more significant digits are given here than justified by the uncertainties, for the sake of reproducibility.) Alternatively, one can view eqs.~(\ref{eq:referencemodelMSbarstart})-(\ref{eq:referencemodelMSbarend}) as the inputs that predict the on-shell quantities in eqs.~(\ref{eq:GFbenchmark})-(\ref{eq:leptonsbenchmark}) as outputs, as can be verified using the utility {\tt calc\_all} of {\tt SMDR} 1.3. Now one can run these parameters to any other renormalization scale $Q$ using the state-of-the-art multiloop Standard Model beta functions, also implemented in {\tt SMDR}. The beta functions for Standard Model parameters were obtained at 2-loop order in refs.~\cite{Jones:1974mm,Caswell:1974gg,Jones:1981we,Machacek:1983tz,Machacek:1983fi,Machacek:1984zw} and \cite{Ford:1992pn,Luo:2002ey}. Also included in {\tt SMDR} are the gauge coupling beta functions through 3-loop \cite{Tarasov:1980au,Larin:1993tp,Mihaila:2012fm,Mihaila:2012pz,Bednyakov:2012rb}
and 4-loop \cite{vanRitbergen:1997va,Czakon:2004bu,Bednyakov:2015ooa,Bednyakov:2016uia,Zoller:2015tha,Poole:2019txl,Davies:2019onf}
orders and the QCD coupling at 5-loop order \cite{Baikov:2016tgj,Herzog:2017ohr,Luthe:2017ttg}, the Yukawa coupling beta functions through 3-loop \cite{Chetyrkin:2012rz,Bednyakov:2012en,Bednyakov:2014pia} order with 
QCD corrections at 4-loop  \cite{Chetyrkin:1997dh,Vermaseren:1997fq}
and 5-loop \cite{Baikov:2014qja} orders, and the 
beta functions for $\lambda$, $m^2$, and $v^2$ through 3-loop order \cite{Chetyrkin:2012rz,Chetyrkin:2013wya,Bednyakov:2013eba}, with 4-loop corrections for the beta function of $\lambda$ at leading order in QCD \cite{Martin:2015eia,Chetyrkin:2016ruf}.
Then, at the new renormalization scale $Q$, one can recompute $G_F$, as a test of the residual scale dependence, putting a lower bound on the theoretical error due to the neglect of higher-order effects. Note that the beta functions incorporate more effects than the present $G_F$ calculation, 
including effects of electroweak gauge couplings at 3-loop and 4-loop order, and QCD effects at 5-loop order.

\begin{figure}[!t]
\begin{minipage}[]{0.6\linewidth}
\begin{flushleft}
\includegraphics[width=0.97\linewidth]{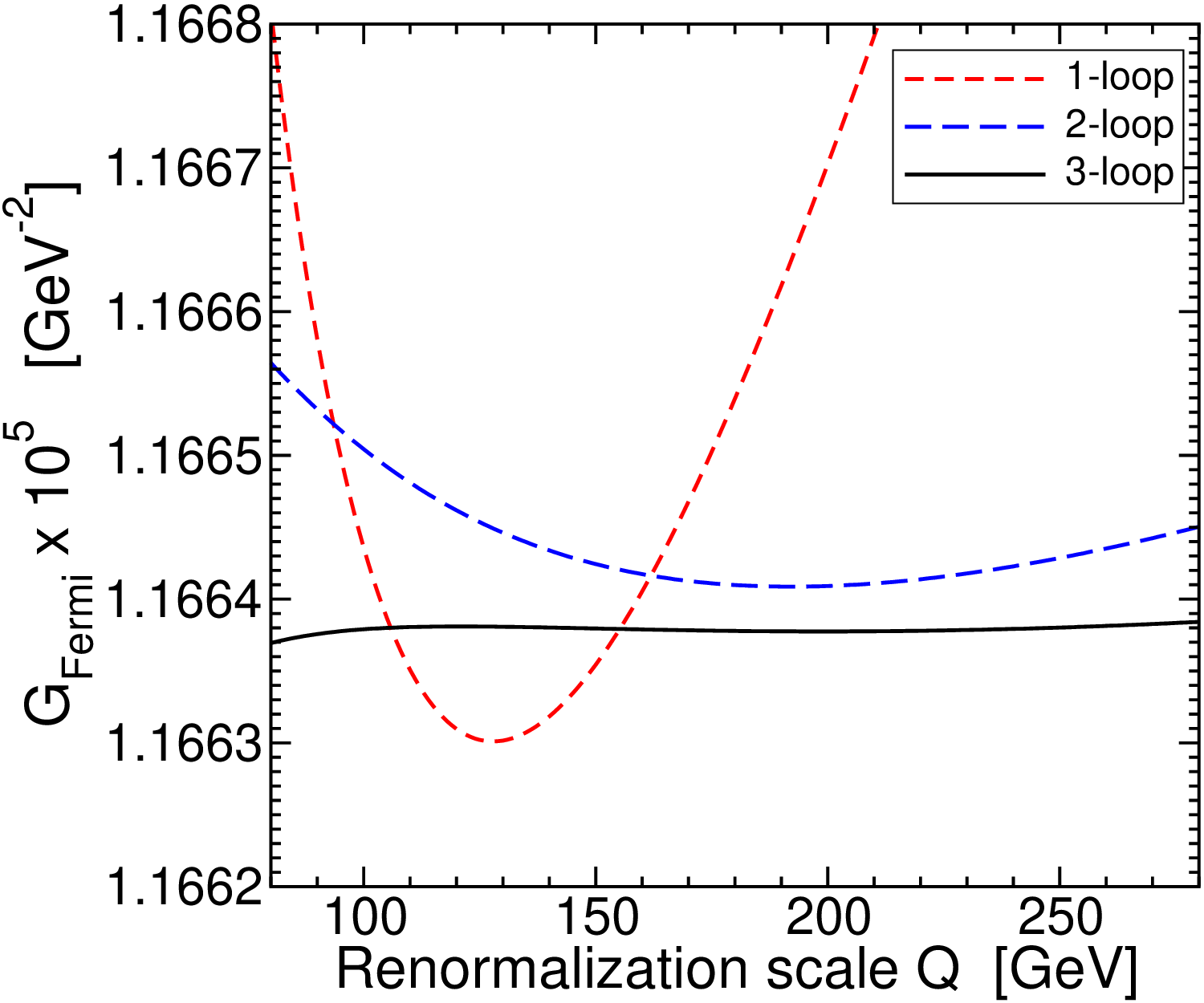}
\end{flushleft}
\end{minipage}
\begin{minipage}[]{0.39\linewidth}
\caption{\label{fig:GFermi} The Fermi decay constant $G_F$, as a function of
the renormalization scale $Q$ at which it is computed from the \MSbar input parameters obtained by
running from the benchmark point at $Q = 172.4$ GeV in eq.~(\ref{eq:referencemodelMSbarstart})-(\ref{eq:referencemodelMSbarend}). These parameters were chosen to yield the on-shell parameters in eqs.~(\ref{eq:GFbenchmark})-(\ref{eq:leptonsbenchmark}).
The short-dashed, long-dashed, and solid lines show the results of including the 1-loop, 2-loop, 
and the 3-loop contributions obtained in the present paper, respectively. The short C program used to make the data plotted in this figure will be included with the v1.3 release of the program {\tt SMDR} \cite{Martin:2019lqd}.}
\end{minipage}
\end{figure}

The results are shown in Figure \ref{fig:GFermi}.
%%%%%%%%%%%%%%%%%%%%%%%%%%%%%%%%%%%%%%%%%%%%%%%%%%%%%%%%%%%%%%%%%%%%%%%%%%%%%%%%%
%%%%%%%%%%%%%%%%%%%%%%%%%%%%%%%%%%%%%%%%%%%%%%%%%%%%%%%%%%%%%%%%%%%%%%%%%%%%%%%%%
The observed scale dependence is greatly reduced by the inclusion of the $\Delta \tilde r^{(3)}$
contributions found in the present paper. I find that for the range of scales $Q$ between 90 GeV 
and 250 GeV, the computed value of $G_F$ in the \MSbar scheme varies by less than a factor of
$\pm 2 \times 10^{-6}$ from its benchmark value at $Q = 172.4$ GeV (the point at which the parameters were fixed in order to make the computed $G_F$ exactly equal to its experimental value). 
The short C program {\tt fig\_GFermi\_vs\_Q.c} that produces the data plotted in this figure will be included with the v1.3 release of the program {\tt SMDR} \cite{Martin:2019lqd}.

%%%%%%%%%%%%%%%%%%%%%%%%%%%%%%%%%%%%%%%%%%%%%%%%%%%%%%%%%%%%%%%
\section{Outlook\label{sec:outlook}}
\setcounter{equation}{0}
\setcounter{figure}{0}
\setcounter{table}{0} 
\setcounter{footnote}{1}

The most direct impact of the new result found here is on the estimate of the Standard Model VEV, which for the benchmark parameter set was found in eq.~(\ref{eq:referencemodelMSbarhiggs}).
As a reminder, the precise definition of this $v$ is that it is the minimum of the complete loop-corrected effective potential in Landau gauge at the specified renormalization scale. Numerically, this value for $v$ is about 3.5 MeV lower than the previous result obtained without the 3-loop contributions to $\Delta \tilde r$, for the same on-shell parameters and fixed $Q=172.4$. 
After including the 3-loop effects found here, a naive estimate of the remaining purely theoretical error (based on the \MSbar scale dependence found above) for the VEV is now an order of magnitude smaller, and well under 1 MeV. However, it is important to recognize that  the true theoretical error could be larger than indicated by the residual scale dependence.

For comparison, the parametric uncertainty in the VEV, coming predominantly from the experimental uncertainty in $M_t$ and to a lesser extent $\alpha_S^{(5)}(M_Z)$ and $M_h$, is much larger. For the change in the VEV (at a fixed scale $Q$) due to a change in these parameters, I find approximately
\beq
\Delta v &=& 10.8\>{\rm MeV} \left (\frac{\Delta M_t}{\rm GeV} \right )
-1.2\>{\rm MeV} \left (\frac{\Delta\alpha_S^{(5)}(M_Z)}{0.001}\right ) 
-0.9\>{\rm MeV} \left (\frac{\Delta M_h}{0.2\>\rm GeV} \right ) .
\eeq

As noted in the introduction, one of the primary uses of the calculation of the Fermi constant in the on-shell scheme is to make a prediction for the $W$ boson mass in the Standard Model. In the pure \MSbar scheme, things are organized differently, as the $W$ and $Z$ boson pole masses are obtained in separate calculations, with the state-of-the art results found in 
refs.~\cite{Martin:2015lxa,Martin:2015rea,Martin:2022qiv} and implemented in {\tt SMDR}. 
For a fixed value of $M_Z$, I find that the effect of including $\Delta\tilde r^{(3)}$ in the fit is to raise $M_W$ by only about $0.5$ MeV.
As found in ref.~\cite{Martin:2022qiv}, for fixed $M_Z$ and other on-shell quantities, the pure \MSbar prediction for $M_W$  is between
the on-shell scheme (higher) \cite{Awramik:2003rn} and hybrid scheme (lower) \cite{Degrassi:2014sxa} predictions, and is consistent with an estimate of a several MeV purely theoretical uncertainty for all three approaches. 
To improve the theoretical situation for the $M_W$ prediction, it would be useful to include the missing (non-leading) 3-loop contributions in all three approaches. However, for the foreseeable future, the parametric uncertainties from the experimental top-quark and $Z$-boson masses, the QCD coupling, and  $\Delta \alpha_{\rm had}^{(5)}(M_Z)$ are responsible for the greater part of the uncertainty in the Standard Model prediction for $M_W$.
 
Acknowledgments: This work is supported in part by the National Science Foundation grant with award number 2310533.

%%%%%%%%%%%%%%%%%%%%%%%%%%%%%%%%%%%%%%%%%%%%%%%%%%%%%%%%%%%%%%%%%%%%

\end{document}